# Growth and transport properties of Mg$_3$$X$$_2$ ($X$ = Sb, Bi) single crystals

Jiazhan Xin[a,b], Guowei Li[a], Gudrun Auffermann[a], Horst Borrmann[a], Walter Schnelle[a], Johannes Gooth[a], Xinbing Zhao[b,*], Tiejun Zhu[b,*], Claudia Felser[a], Chenguang Fu[a,*]

[a] Max Planck Institute for Chemical Physics of Solids, Nöthnitzer Str. 40, 01187 Dresden, Germany

[b] State Key Laboratory of Silicon Materials, School of Materials Science and Engineering, Zhejiang University, 310027 Hangzhou, China

*Corresponding author.
E-mail addresses*: chenguang.fu@cpfs.mpg.de (C.G. Fu); zhaoxb@zju.edu.cn (X.B. Zhao); zhutj@zju.edu.cn (T.J. Zhu)

## Abstract

The discovery of high thermoelectric performance in *n*-type polycrystalline Mg$_3$(Sb,Bi)$_2$ based Zintl compounds has ignited intensive research interest. However, some fundamental questions concerning the anisotropic transport properties and the origin of intrinsically low thermal conductivity are still elusive, requiring the investigation of single crystals. In this work, high-quality *p*-type Mg$_3$Sb$_2$ and Mg$_3$Bi$_2$ single crystals have been grown by using a self-flux method. The electrical resistivity $\rho$ of Mg$_3$Bi$_2$ single crystal displays an anisotropy with $\rho$ in-plane twice larger than out-of-plane. The low-temperature heat capacity and lattice thermal conductivity of Mg$_3$Sb$_2$ and Mg$_3$Bi$_2$ single crystals have been investigated by using the Debye-Callaway model, from which the existence of low-lying vibration mode could be concluded. Large Grüneisen parameters and strong anharmonicity are found responsible for the intrinsically low thermal conductivity. Moreover, grain boundary scattering does not contribute significantly to suppress the lattice thermal conductivity of polycrystalline Mg$_3$Sb$_2$. Our results provide insights into the intrinsic transport properties of $Mg_3X_2$ and could pave a way to realize enhanced thermoelectric performance in single-crystalline $Mg_3X_2$-based Zintl compounds.

## 1. Introduction

Solid-state thermoelectric (TE) technology, enabling direct and reversible energy conversion between heat and electricity, offers a promising solution for some of today's energy problems and thus, has recently attracted considerable research attention.[1-3] In the past two decades, under tremendous and continuous exploratory efforts, more than ten material systems have been developed as high-performance TE materials with the figure-of-merit $zT$ value exceeding unity, such as bismuth and lead chalcogenides,[4-6] tin selenides,[7,8] copper chalcogenides,[9,10] filled skutterudites,[11-13] half-Heusler alloys,[14-17] and Zintl compounds,[18,19] *etc*. In the perspective of practical application, good TE materials with earth-abundant, low-cost and non-toxic elements are of preference. Among them, Zintl compounds are one of the families possessing these good characteristics. Recently, the surprising discovery of high-performance in *n*-type $Mg_3(Sb,Bi)_2$-based Zintl compounds has ignited intensive research activity on this system,[20, 21] especially since almost all the other known Zintl compounds display only good *p*-type TE properties due to their intrinsic point defects.[22, 23]

The Zintl phase $Mg_3Sb_2$, which crystalizes in a hexagonal layered structure composed of $Mg^{2+}$ and $(Mg_2Sb_2)^{2-}$ layers, was discovered for the first time in 1933.[24] Since the beginning of this century, research attention has been paid to this compound but initial studies only focused on its *p*-type TE properties.[25-27] In 2016, Tamaki *et al.* reported a prominent $zT$ of 1.5 at 716 K in *n*-type $Mg_{3.2}Sb_{1.5}Bi_{0.49}Te_{0.01}$, and this result has soon been verified by independent groups.[20, 21] Significant features contributing to the excellent TE performance have subsequently been identified. The multi-valley band behavior combined with light band effective mass, yielding large TE quality factor $\beta$, were found to be important reasons underlying the high $zT$.[21] The diminutive size of Mg in $Mg_3Sb_2$, leading to weak interlayer bonding, was attributed to be responsible for its low thermal conductivity.[28] More attempts have been made to further optimize the TE performance of the $Mg_3Sb_2$-based compounds, such as proper selection of the dopants on the anion site,[29] rational choice of the alloying compositions (alloyed with $Mg_3Bi_2$),[30] manipulation of the carrier scattering mechanisms[31, 32] and utilization of the possible anisotropic transport characteristics (forming textures)[33].

The above-mentioned experimental works are all focused on polycrystalline samples while some fundamental questions related to the intrinsic physical properties of this material system are still elusive. For instance, the observed increasing trend of electrical conductivity ($\sigma$) with increasing temperature ($T$) in both $p$-type and $n$-type $Mg_3Sb_2$ at around room temperature was intuitively regarded as the results of ionized impurity scattering on carriers.[20, 34] But it was more recently found that such temperature dependence of carrier mobility ($\mu$) could be altered by enlarging the grain size, indicating that grain boundary scattering might dominate electron transport.[32] Theoretical modeling on the possible grain boundary scattering on carriers was proposed and approximately 60% improvement of $zT$ was predicted at room temperature if grain boundary scattering were eliminated.[35] Furthermore, given the layered structure of $Mg_3(Sb,Bi)_2$, anisotropic electrical and thermal transport properties are expected. Nevertheless, previous studies did not show obvious anisotropy in transport properties along different directions for the polycrystalline $Mg_3Sb_{1.5}Bi_{0.5}$ synthesized by uniaxial-press sintering.[20, 33, 36] To achieve low thermal conductivity ($\kappa$) and high $zT$, previous studies mainly focus on the solid solution composition, while the basic physical properties of the parent compounds are less studied. Interestingly, $Mg_3Bi_2$ was predicted to be a topological nodal-line semimetal by first-principles calculations and subsequently verified by angle-resolved photoemission spectroscopy.[37, 38] However, the magneto-transport properties of this compound are still missing.

In order to address the above-mentioned questions, it is appealing to grow $Mg_3Sb_2$ and $Mg_3Bi_2$ single crystals and to investigate their intrinsic transport properties. In a previous work by Kim *et al.*, $Mg_3Sb_{2-x}Bi_x$ single crystals were grown by using the Bridgman method and the electrical and thermal transport properties below room temperature were reported.[39] Surprisingly, they reported a huge anisotropy in room temperature electrical resistivity $\rho$ of $Mg_3Bi_2$ single crystal with the out-of-plane $\rho$ 100 times larger than the in-plane $\rho$.[39] The in-plane room temperature carrier mobility of $Mg_3Bi_2$ single crystal was reported to be only about 1 $cm^2V^{-1}s^{-1}$.[39] As $Mg_3Bi_2$ was predicted to be a semimetal, such low carrier mobility in its single crystalline form is unexpected.[30, 40] These contradicting findings also motivate us to reinvestigate the intrinsic transport properties of $Mg_3Sb_2$ and $Mg_3Bi_2$ single crystals.

In this work, we have successfully grown $Mg_3X_2$ ($X$ = Sb, Bi) single crystals by using a self-flux method. Electrical transport properties measurement shows that the out-of-plane $\rho$ is about half smaller than the in-plane $\rho$ in $Mg_3Bi_2$ single crystal, which is consistent with the electronic structure calculation showing smaller effective mass along the out-of-plane direction.[17] Near room temperature, the electron transport of $Mg_3Bi_2$ is dominated by acoustic phonon scattering. Heat capacity and lattice thermal conductivity of $Mg_3X_2$ single crystals have been investigated by using the Debye-Callaway model. Large Grüneisen parameter and strong anharmonicity are found responsible for the intrinsically low thermal conductivity. Our results provide further insights into the intrinsic transport properties of $Mg_3X_2$ system.

## 2. Experimental section

The single crystals of $Mg_3Sb_2$ and $Mg_3Bi_2$ were grown by using a self-flux method with Sb or Bi as flux, respectively. Starting elements Mg (granules, 99.8%), Sb (shot, 99.999%) and Bi (shot, 99.999%) were weighed and mixed with molar ration of Mg: Sb/Bi = 3:7. Tantalum tubes with diameter of 10 mm were used to seal the mixtures in an argon glove box. These Ta tubes were then sealed in quartz tubes under partial argon pressure. For $Mg_3Sb_2$, the ampoule was heated up to 850 °C, kept for 24 hours and then slowly cooled down (2 °C/h) to 630 °C. After that, the ampoule was ramped up to 650 °C and the crystals were separated from the Sb flux through a centrifuging process. It was noted that a thin dense layer of $TaSb_2$ was formed on the inner surface of the Ta tube during the heating process due to slight reaction between Sb flux and Ta tube.[41] The single crystals of $Mg_3Sb_2$ tended to "grow" on the $TaSb_2$ layer, making it easier to separate $Mg_3Sb_2$ from Sb flux. The obtained single crystals typically had lamellar shape as seen in Fig. S1 with thickness of 0.2 - 1.5 mm, and the flat face was identified as the $ab$-plane with typical length up to 6 - 7 mm. For $Mg_3Bi_2$, the mixture was heated up to 650 °C, kept for 24 hours and then slowly cooled down (1.5 °C /h) to 350 °C. Then, the single crystals were separated from the Bi flux through a centrifuging process. $Mg_3Bi_2$ single crystals exhibit some plasticity compared to the brittle $Mg_3Sb_2$ single crystals.

The chemical compositions of the single crystals were checked by using a scanning electron microscope (SEM) with an attached energy-dispersive X-ray spectrometer (EDX), which are in agreement with the nominal compositions. The crystals were checked and oriented at room temperature by using a four-circle Rigaku AFC7 X-ray diffractometer with Saturn 724+ CCD detector. Suitable sample edges were selected where transmission of $Mo\text{-}K_\alpha$ ($\lambda$ = 0.71073 Å) radiation seemed feasible. After successful indexing, oscillation images about the crystallographic axis allowed the assignment of the crystal orientation, confirmed the appropriate choice of the unit cell and showed the excellent crystal quality. The single crystals were then cut along in-plane $b$-axis (010) and out-of-plane $c$-axis (001) directions by using a wire saw for the measurements of anisotropic transport properties. The single crystal nature of the as-cutted $Mg_3Sb_2$ crystal was further checked by white-beam backscattering Laue X-ray diffraction method. Powder XRD measurements were also performed to check the crystal structure and phase purity, with Cu $K_\alpha$ radiation, and using an image-plate Huber G670 Guinier camera equipped with a Ge (111) monochromator.

Both longitudinal and Hall resistivities of $Mg_3Bi_2$ single crystals were measured by a standard four-probe method using the AC transport option in a PPMS (Quantum

Design). For Mg$_3$Sb$_2$ single crystal, since its resistance is too high, a home-made setup was employed for the electrical transport measurement. As resistivity displays an exponential increase with decreasing temperature, it is difficult to make accurate measurement at low temperature. All contacts were made by using silver-filled paint and 25 μm Platinum wire. To correct for contact misalignment, the measured Hall resistivity was field anti-symmetrized. The Hall coefficient $R_H$ was determined from isothermal magnetic field dependent Hall resistivities ($\rho_H$). The charge carrier concentration $n_H$ was calculated by $n_H = 1/eR_H$, where $e$ is the unit charge. The thermal conductivity and Seebeck coefficient ($S$) of Mg$_3$Bi$_2$ and Mg$_3$Sb$_2$ were measured along $b$ axis by the one-heater and two-thermometer configuration using the thermal transport option (TTO) of the PPMS. The contact was made by using silver-filled epoxy and copper flat wire. The applied temperature gradients were about 1 % of the base temperature. The Seebeck coefficient of Mg$_3$Sb$_2$ could not be accurately measured due to extremely high electrical resistance of the sample. The heat capacity ($C_p$) of Mg$_3$Sb$_2$ and Mg$_3$Bi$_2$ single crystals was measured from 1.9 K to 320 K by using the heat capacity option of the PPMS.

## 3. Results and Discussion

The synthesized single crystal of Mg$_3$Sb$_2$ displays shiny surface and sharp edges. Clean surface is observed from the SEM image of Mg$_3$Sb$_2$, as displayed in Fig. 1a. Powder XRD pattern of Mg$_3$Sb$_2$ shows no obvious impurity phase (Fig. S2, Supporting Information SI), indicating good phase purity of the single crystal. For Mg$_3$Bi$_2$, large surfaces of the crystal are clean and flat, but some remaining Bi flux on the sample surface is also found in the SEM image (Fig. 1d). The reflection pattern of Bi is also detectable in the powder XRD pattern of Mg$_3$Bi$_2$ (Fig. S2b), which is probably due to the inadequate centrifuging process. Thus, before performing the transport measurement, the crystal was well polished to remove the Bi flux. The calculated lattice parameters of both crystals are shown in Fig. S2 and are in agreement with the literature data.[42, 43] Furthermore, single crystal XRD diffraction images of $Mg_3Sb_2$ and $Mg_3Bi_2$ along *b* and *c* directions respectively are shown in Fig. 1(b-c, e-f). Sharp and distinct spots are observed, which indicates good quality of the crystals. The faint powder rings in Fig.1e-f are probably attributed to the slight distortions or powder contaminations on the surface. More information about single crystal XRD results can be found in Fig. S3-S4. Laue diffraction pattern of $Mg_3Sb_2$ crystal is further shown in Fig. S5.

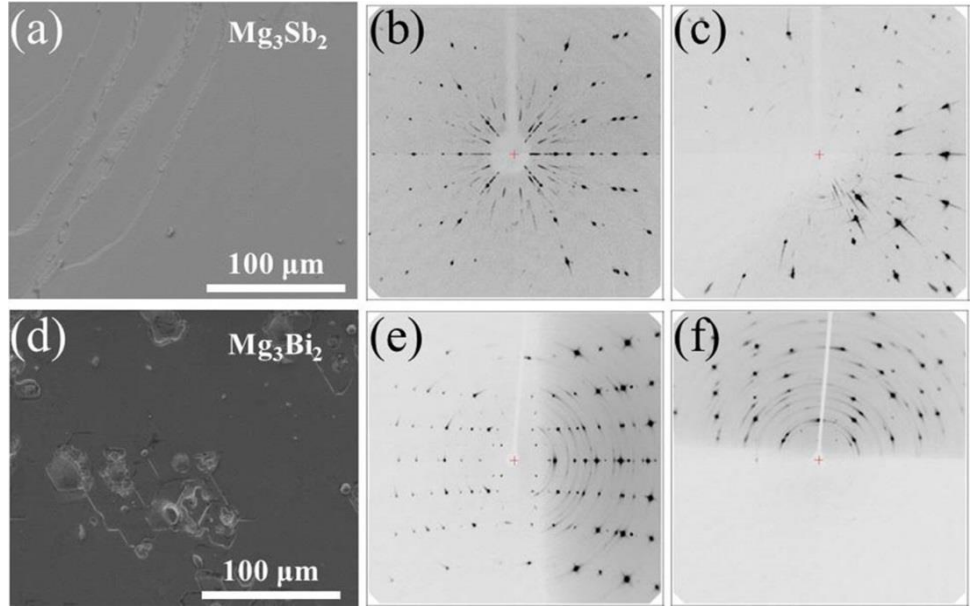

**Fig. 1.** SEM and single crystal diffraction images for selected crystals of $Mg_3Sb_2$ (a - c) and $Mg_3Bi_2$ (d – f). (b) and (e) are reflecting oscillations about *b*-axis, (c) and (f) about *c*-axis. Reflection broadening is due to surface distortion induced by sample preparation.

$Mg_3X_2$ crystallize in the hexagonal anti-$La_2O_3$ structure type with a cationic $Mg^{2+}$ sheet and an anionic $[Mg_2X_2]^{2-}$ layer stack along the *c* axis (Fig. 2a). This layered structure raises the question whether the electrical and thermal transport properties of $Mg_3(Sb,Bi)_2$ are anisotropic, *i.e.*, different for in-plane (*ab* plane) and out-of-plane (*c* axis) directions. Tamaki *et al.* calculated the sound velocities of the longitudinal acoustic mode of $Mg_3Sb_2$ along ΓM direction (in-plane) and ΓA (out-of-plane) and they found similar values along these two directions and thus, concluded that the thermal conduction should be isotropic.[20] The electronic structure of $Mg_3X_2$ has been calculated by Zhang *et al*, whose results show that the Fermi surfaces in conduction band minimum are approximately spherical while in valence band maximum the in-plane effective mass is about 10 times larger than the out-of-plane one, indicating isotropic and anisotropic electrical transport properties in *n*-type and *p*-type $Mg_3X_2$, respectively.[21] In experiments, the TE properties of *n*-type polycrystalline $Mg_3(Sb,Bi)_2$ along directions parallel and perpendicular to the pressing direction was reported, showing no obvious difference.[33]

To give our answer to the question of anisotropy, we measured the temperature-dependent electrical resistivity of the $Mg_3Bi_2$ and $Mg_3Sb_2$ single crystals along the in-plane *b* axis and the out-of-plane *c* axis, respectively. As both crystals were grown in the Mg-poor chemical environment, they exhibit *p*-type conducting behavior. The results, which would reflect the anisotropy of the valence band, are displayed in Fig. 2b and 2c. Distinct from the previous report by Kim *et al.*, who found a huge anisotropy in $Mg_3Bi_2$ with the out-of-plane resistivity 100 times larger than the in-plane resistivity,[39] our results show the resistivity along out-of-plane *c* axis is approximately half of that along in-plane *b* axis in the whole investigated temperature range. While both Kim's and our crystals display positive Seebeck coefficient (seen in Fig. S6), indicating the electrical transport is dominated by holes, the huge difference in the anisotropy is confusing. Recalling the electronic structure calculation by Zhang *et al.*,[21] showing lower valence band effective mass along *c* axis, our finding of lower resistivity along *c* axis is in qualitative agreement with the expecting result from band structure. The resistivity of $Mg_3Sb_2$ is about 6 orders of magnitude higher than that of $Mg_3Bi_2$ and shows typical semiconducting behavior, which indicates that the Fermi level should lie in the forbidden gap. The anisotropic resistivity is more complicated. As shown in Fig. 2c, $\rho_c$ is smaller than $\rho_b$ at around room temperature, whereas $\rho_c$ becomes larger than $\rho_b$ below 255K. Thus, the difference in $\rho_c$ and $\rho_b$ of $Mg_3Sb_2$ might not give accurate information concerning the anisotropy of electrical transport properties.

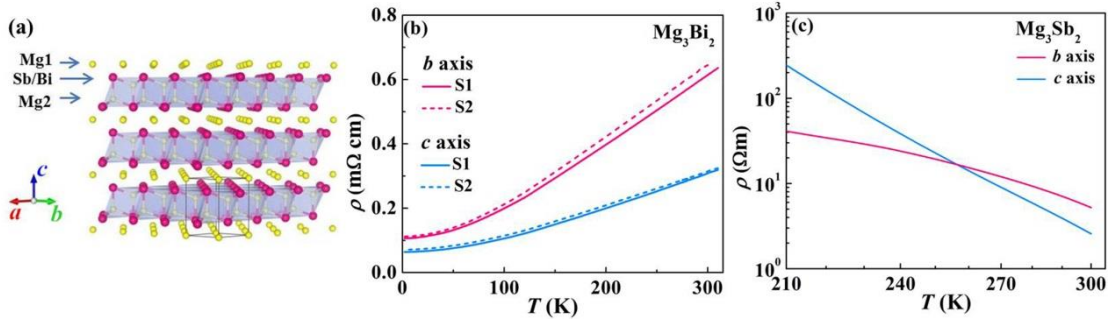

**Fig. 2.** (a) Crystal structure of $Mg_3Bi_2$ and $Mg_3Sb_2$ (drawn using VESTA)[44]; Temperature dependent electrical resistivity of (b) $Mg_3Bi_2$ and (c) $Mg_3Sb_2$ single crystals along in-plane *b* axis and out-of-plane *c* axis, respectively. To confirm the result of anisotropic resistivity, two different $Mg_3Bi_2$ single crystals, denoted as S1 and S2 respectively, were selected from the same batch for the measurement, which give consistent results.

$Mg_3Bi_2$ was predicted to be a topological type-II nodal-line semimetal by first-principles calculations and subsequently verified by angle-resolved photoemission spectroscopy, but its magneto-transport properties were not reported yet.[37, 38] Many discovered topological semimetals show large magnetoresistance (MR) and high carrier mobility, which motivates us to investigate the MR of $Mg_3Bi_2$ single crystal. As shown in Fig. 3a, the MR displays a nearly linear behavior with the increase of magnetic field in the investigated temperature range. A maximum MR of 3% is obtained at 2 K and 9 T, which however, is much smaller than that of other topological semimetals, such as the Dirac semimetals $Cd_3As_2$ and $PtSn_4$[45, 46] or the Weyl semimetals NbP and TaAs.[47-49] To explore the possible reason leading to this low MR, we further measured the Hall resistivity $\rho_H$ of $Mg_3Bi_2$, as shown in Fig. 3b. The $\rho_H(H)$ shows a positive and linear increase with magnetic field, indicating that the electrical transport properties is dominated by holes. The Hall carrier concentration is further calculated by the slope of $\rho_H(H)$, as displayed in Fig. 3c. A slight increase in $n_H$ with rising temperature is found and the $n_H(T)$ at room temperature is about $2.5 \times 10^{20}$ cm$^{-3}$. The calculated carrier mobility $\mu_H(T)$ decreases with rising temperature. Especially above 100 K, $\mu_H(T)$ follows a $T^{-1}$ relationship, which is a typical behavior for heavily doped semiconductors or metals when the acoustic phonon scattering dominates the electron transport. $\mu_H(T)$ of $Mg_3Bi_2$ at 2 K is around several hundreds of cm$^2$V$^{-1}$s$^{-1}$, which is much smaller than that of the other topological semimetals. For example, the carrier mobility of NbP single crystal at 2 K is almost 4 orders of magnitude higher than that of Mg$_3$Bi$_2$.[47] Thus, the low carrier mobility could be the one of the reasons why Mg$_3$Bi$_2$ has much lower MR compared to other topological semimetals.

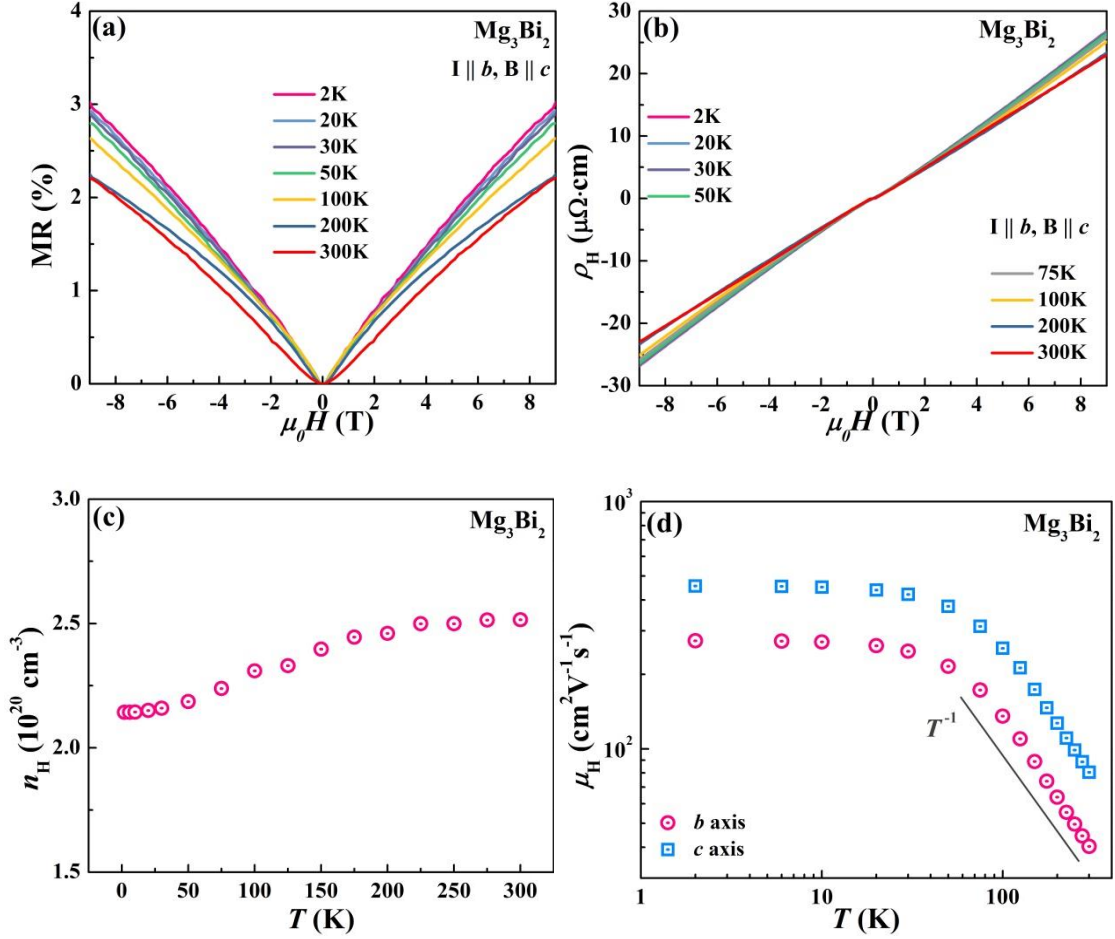

**Fig. 3.** (a) Magnetoresistance (MR) and (b) Hall resistivity of the Mg$_3$Bi$_2$ single crystal at different temperatures measured in magnetic fields up to 9 T. The electrical current was applied along *b* axis and the magnetic field along *c* axis. (c) Temperature-dependent Hall carrier concentration $n_H$ and (d) carrier mobility $\mu_H$. The positive slop of Hall resistivity indicates *p*-type conducting behavior.

The thermal conductivity and lattice thermal conductivity of single crystalline Mg$_3$Sb$_2$ and Mg$_3$Bi$_2$ along *b* axis are shown in Fig. 4. The lattice thermal conductivity of Mg$_3$Bi$_2$ was calculated by subtracting the electronic term $\kappa_e$ which was determined according to the Wiedemann–Franz law, $\kappa_e = L\sigma T$, using the temperature dependent Lorenz number $L$, which was roughly calculated from the Seebeck coefficient based on a single parabolic band model.[50] The lattice thermal conductivity of $Mg_3Sb_2$ is almost the same with its thermal conductivity due to its extremely high electrical resistivity and thus negligible electronic thermal conductivity. The lattice thermal conductivity of both $Mg_3Sb_2$ and $Mg_3Bi_2$ exhibits normal temperature dependency, in which $\kappa_L$ first increases with increasing temperature, reaches its peak value at around $\theta_D/20$ ($\theta_D$ is 241 K and 175 K respectively, as shown below), and then decreases at higher temperature. Above 20 K, the $\kappa_L(T)$ of $Mg_3Sb_2$ exhibits a nearly $T^{-1}$ relationship due the dominant phonon-phonon Umklapp scattering.

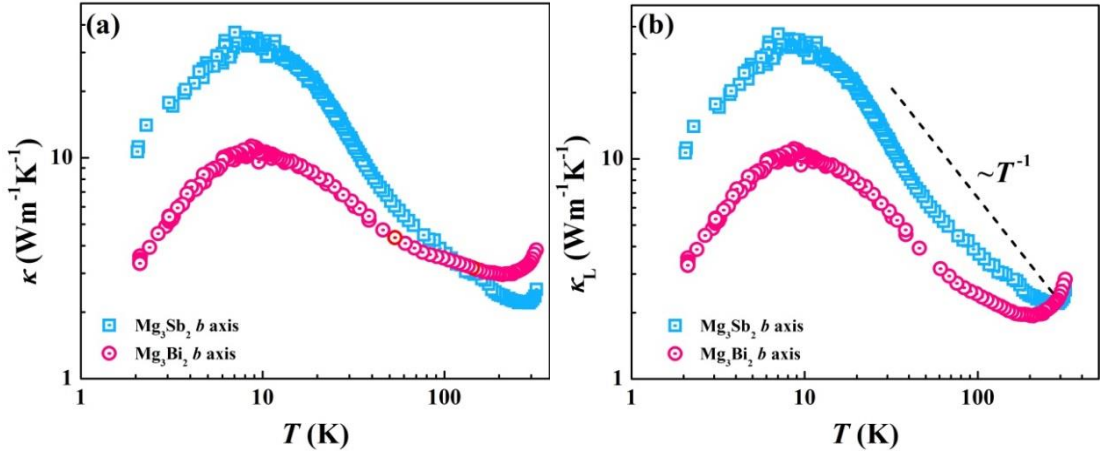

**Fig. 4.** Temperature dependences of (a) thermal conductivity and (b) lattice thermal conductivity of the $Mg_3Sb_2$ and $Mg_3Bi_2$ single crystals along $b$ axis.

One interesting question is why $Mg_3Sb_2$ and $Mg_3Bi_2$ have lower $\kappa_L$ compared to the other isostructural $AMg_2X_2$ ($A$ = Mg, Ca, and Yb) Zintl compounds like $CaMg_2Sb_2$ and $CaMg_2Bi_2$. This issue has recently been discussed by Peng *et al.*[28] They proposed that the diminutive size of $Mg^{2+}$ leads to low-lying transverse phonon modes and large mode Grüneisen parameters, which thus contribute to the inherently low $\kappa_L$ of $Mg_3X_2$. To check for the existence of low-lying vibration modes,[28] the low-temperature heat capacity of both $Mg_3Sb_2$ and $Mg_3Bi_2$ was further measured and is shown in Fig. 5a. As only the low frequency phonons could be excited at $T \ll \theta_D/10$ (long wave length limit), the $C_p$ at that temperatures would obey the Debye $T^3$ law as seen in Fig. 5b. Therefore, the Debye temperatures of these two compounds were calculated by using equation $C_p/T = \delta + 2.4\pi^4 n N_A k_B \theta_D^{-3} T^2$, where $\delta$ is a constant from the contribution of electrons, $N_A$ is the Avogadro constant, $n$ is the number of atoms per primitive cell, $k_B$ is the Boltzmann constant.[51] The calculated $\theta_D$ (241 K for $Mg_3Sb_2$ and 175 K for $Mg_3Bi_2$) is in accordance with the values obtained from theoretical calculations (238 K for $Mg_3Sb_2$ and 177 K for $Mg_3Bi_2$).[52] At higher temperature, the experimental $C_p$ gradually exhibits deviation from the calculated $C_p$ if only based on the Debye $T^3$ approximation, as seen in Fig. 5b. Considering the possible existence of low-lying vibration modes, the Debye-Einstein model, in which additional Einstein oscillators are included, was used to fit the experimental $C_p$. The additional Einstein term is expressed as $C_p = A_i(\theta_{Ei})^2 T^{-3}(e^{\theta_{Ei}/T})/(e^{\theta_{Ei}/T}-1)^2$, where $\theta_{Ei}$ is the Einstein temperature for the $i^{th}$ Einstein oscillator and $A_i$ is the fitting parameter (detailed calculation can be found in SI).[53] As seen in Fig. 5c and 5d, one Einstein oscillator is necessary to fit the experimental $C_p$, with Einstein temperature $\theta_{E1}$ of 64 K for $Mg_3Sb_2$ and 44 K for $Mg_3Bi_2$. The corresponding frequency of the Einstein oscillator for $Mg_3Sb_2$ and $Mg_3Bi_2$ is 44.2 cm$^{-1}$ and 30.8 cm$^{-1}$. The derived low-lying vibration mode in $Mg_3Sb_2$ is qualitatively agreeing with the calculation by Peng et al.,[28] showing a low-lying transverse phonon mode with similar frequency and very large Grüneisen parameter at $A$ point of the Brillouin zone. Thus, we could conclude that the existence of low-lying vibration mode in $Mg_3X_2$ system.

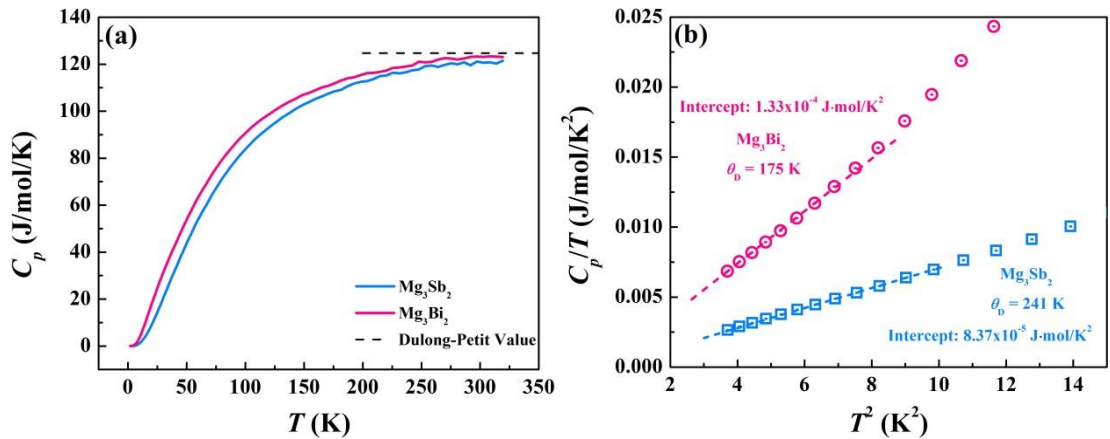

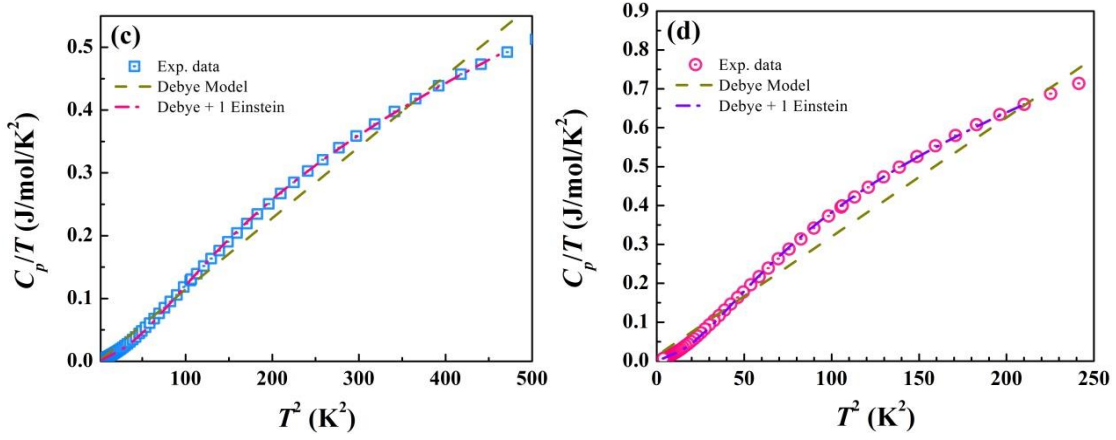

**Fig. 5.** Temperature dependence of specific heat capacity of $Mg_3Sb_2$ and $Mg_3Bi_2$ single crystals, presented as (a) $C_p$ versus $T$ and (b-d) $C_p/T$ versus $T^2$, together with the calculated curves based on Debye-Einstein model (more detailed fitting results could be seen in Fig. S7).

The above analysis of low-temperature $C_p$ shows the existence of low-lying vibration mode, and then it is interesting to quantitatively estimate its influence on the $\kappa_L$ of $Mg_3Sb_2$ and $Mg_3Bi_2$. Herein, the low-lying vibration mode is taken as resonant scattering term, as the other researchers did for the other TE materials.[53, 54] The modified Debye-Callaway model [55] was used to fit the experimental $\kappa_L$, with phonon-phonon scattering (U and N), boundary scattering (B), point defect scattering (PD), resonant scattering (PR) taken into consideration (calculation details seen in SI). The resonant frequency used in calculating the scattering rate of resonant scattering was derived from the fitting results of low temperature specific heat capacity. As seen in Fig. 6, the experimental $\kappa_L$ is in agreement with the theoretical fit by introducing all the scattering processes mentioned above. It is found that the low-lying vibration mode and the induced phonon resonant scattering have strong influence on the $\kappa_L$ within the low temperature region, which is similar to the cases in $\alpha$-MgAgSb and filled skutterudites.[53, 56] Furthermore, the Grüneisen parameters for longitude and transverse phonon branches ($\gamma_L$ and $\gamma_T$) used to fit the experimental data are extremely large ($\gamma_L = 4.33$, $\gamma_T = 2.22$ for $Mg_3Sb_2$ and $\gamma_L = 6.49$, $\gamma_T = 2.22$ for $Mg_3Bi_2$), which is in accordance with the calculated large mode Grüneisen parameters by Peng *et al*.[28] Therefore, we conclude that the large Grüneisen parameter and the strong anharmonicity are the origin of the intrinsically low thermal conductivity of $Mg_3X_2$.

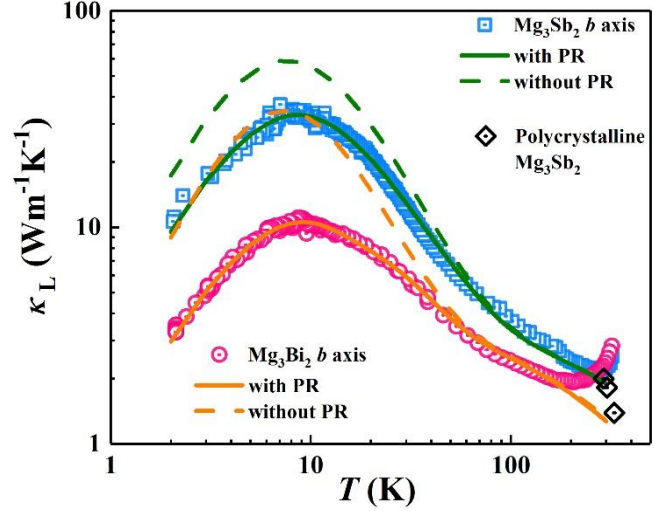

**Fig.6** Temperature dependence of lattice thermal conductivity of single crystalline $Mg_3Sb_2$ and $Mg_3Bi_2$. The solid and dash curves are calculated based on the modified Debye-Callaway model with and without resonant scattering (PR), respectively. The data of polycrystalline $Mg_3Sb_2$ are taken from the references.[20, 26, 34]

As a further discussion, it is worth to roughly estimate the TE performance of single crystalline $Mg_3X_2$ system. The room temperature $\kappa_L$ of polycrystalline $Mg_3Sb_2$ as shown in Fig. 6,[20, 26, 34] is only about 15% lower than that of the single crystalline $Mg_3Sb_2$. This surprisingly small difference in $\kappa_L$ between polycrystalline and single crystalline $Mg_3Sb_2$ indicates that grain boundary scattering does not contribute significantly to further suppress the lattice thermal conductivity of polycrystalline samples. In contrast, since the electron mobility could be largely improved by eliminating the grain boundary, which has been justified in previous work,[32, 35] one can optimistically expect higher TE performance in *n*-type single crystalline $Mg_3X_2$ system.

## 4. Conclusions

In summary, single crystals of $Mg_3Sb_2$ and $Mg_3Bi_2$ have been successfully grown by using the self-flux method. In the Mg-poor synthesis environment, all the obtained single crystals exhibit *p*-type transport properties, of which the $Mg_3Sb_2$ shows typical semiconducting behavior and the $Mg_3Bi_2$ shows semimetal behavior. We found that the electrical resistivity of $Mg_3Bi_2$ single crystal displays an obvious anisotropy with the out-of-plane resistivity about half of the in-plane resistivity. Magnetoresistance of Mg$_3$Bi$_2$ was measured, which shows a maximum value of 3% at 2 K and 9 T. The low-temperature heat capacity and lattice thermal conductivity of Mg$_3$Sb$_2$ and Mg$_3$Bi$_2$ single crystals have been investigated based on the modified Debye-Callaway model, which justifies the existence of low-lying vibration mode. Large Grüneisen parameters and strong anharmonicity were found responsible for the intrinsically low thermal conductivity. Moreover, the lattice thermal conductivity of single crystalline Mg$_3$Sb$_2$ is only about 15% higher than that of the polycrystalline samples, indicating higher TE performance might be achieved in *n*-type single crystalline Mg$_3$X$_2$ system as the carrier mobility can be significantly enhanced.

## Acknowledgements

This work was financially supported by the ERC Advanced Grant No. (742068) "TOP-MAT" and funded by the Deutsche Forschungsgemeinschaft (DFG, German Research Foundation) – Projektnummer (392228380). C. G. Fu acknowledges the financial support from Alexander von Humboldt Foundation. The authors thank Jiong Yang from Shanghai University for helpful discussion.

# Supporting Information

**Growth and transport properties of Mg$_3$$X$$_2$ ($X$ = Sb, Bi) single crystals**

Jiazhan Xin[a,b], Guowei Li[a], Gudrun Auffermann[a], Horst Borrmann[a], Walter Schnelle[a], Johannes Gooth[a], Xinbing Zhao[b,*], Tiejun Zhu[b,*], Claudia Felser[a], Chenguang Fu[a,*]

[a] Max Planck Institute for Chemical Physics of Solids, Nöthnitzer Str. 40, 01187 Dresden, Germany

[b] State Key Laboratory of Silicon Materials, School of Materials Science and Engineering, Zhejiang University, 310027 Hangzhou, China

*Corresponding author.
E-mail: chenguang.fu@cpfs.mpg.de (C. G. Fu); zhaoxb@zju.edu.cn (X. B. Zhao); zhutj@zju.edu.cn (T. J. Zhu)

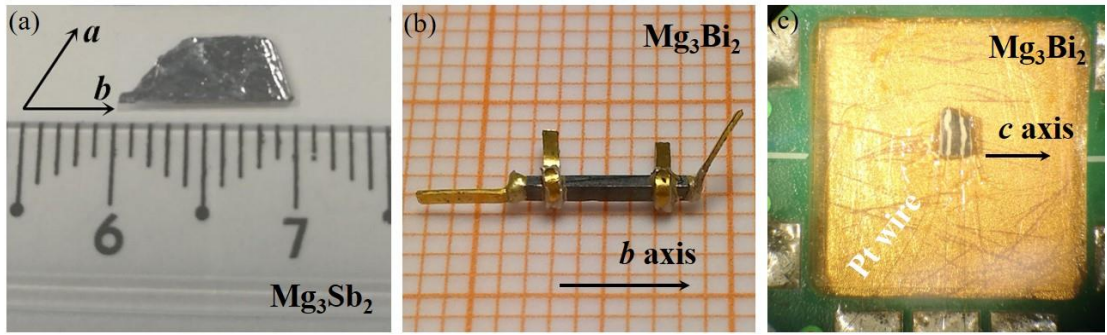

**Fig. S1.** Optical images of the single crystals of (a) $Mg_3Sb_2$ and (b, c) $Mg_3Bi_2$. (b) Four copper leads were used for thermal transport measurement. (c) A typical four-probe electrical resistivity measurement setup for $Mg_3Bi_2$ crystal along $c$-axis with the length of about 0.6 mm.

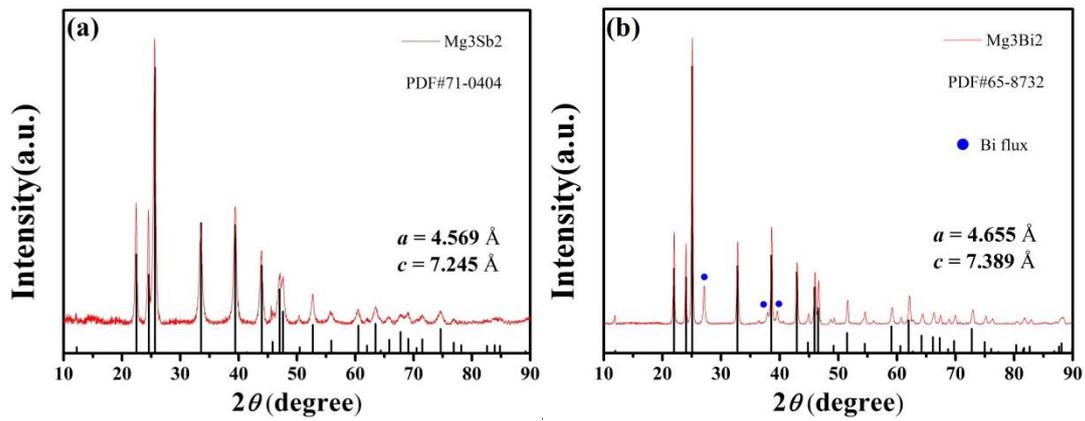

**Fig. S2.** Powder XRD results of (a) $Mg_3Sb_2$ and (b) $Mg_3Bi_2$.

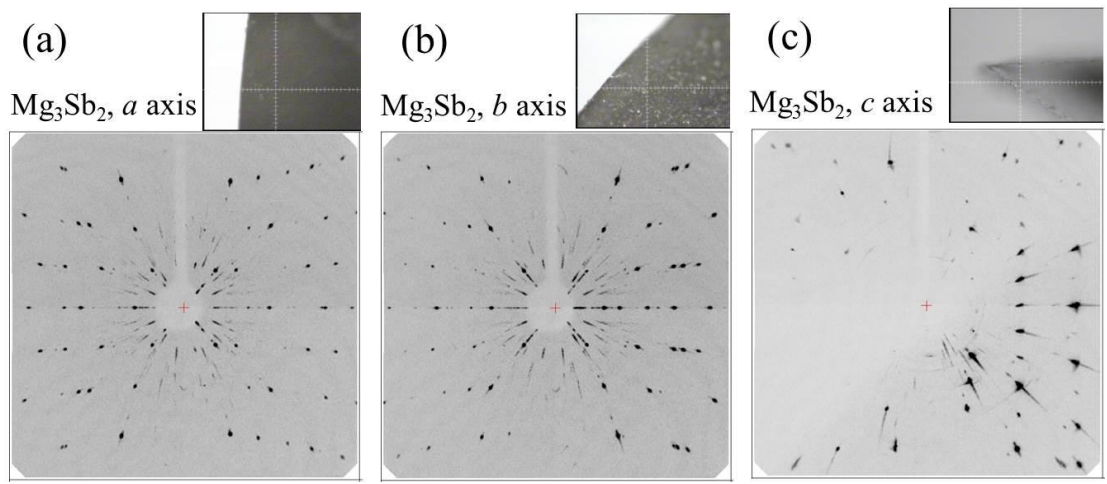

**Fig. S3.** Single crystal XRD results of $Mg_3Sb_2$; the patterns were obtained with the X-ray along (a) *a* axis, (b) *b* axis and (c) *c* axis.

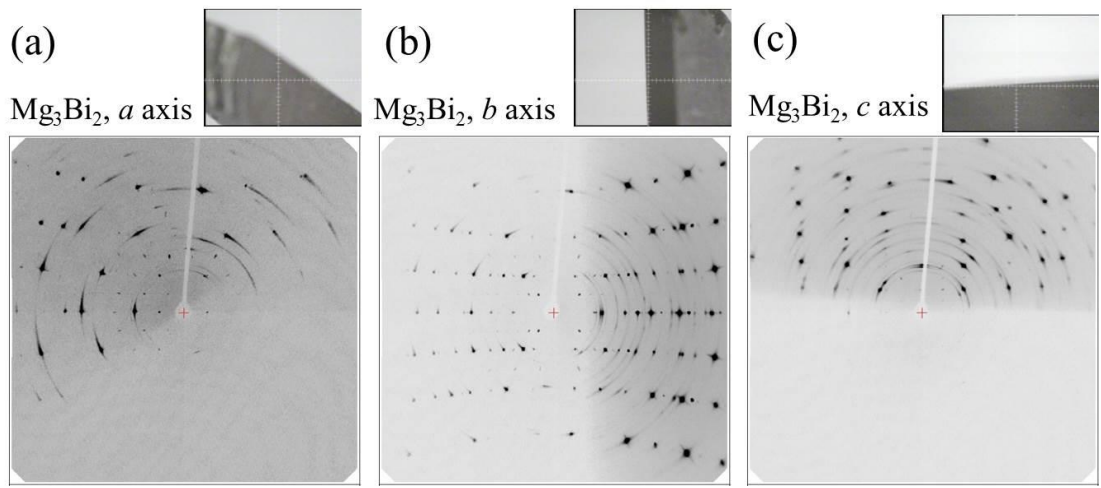

**Fig. S4.** Single crystal XRD results of $Mg_3Bi_2$; the patterns were obtained with the X-ray along (a) *a* axis, (b) *b* axis and (c) *c* axis.

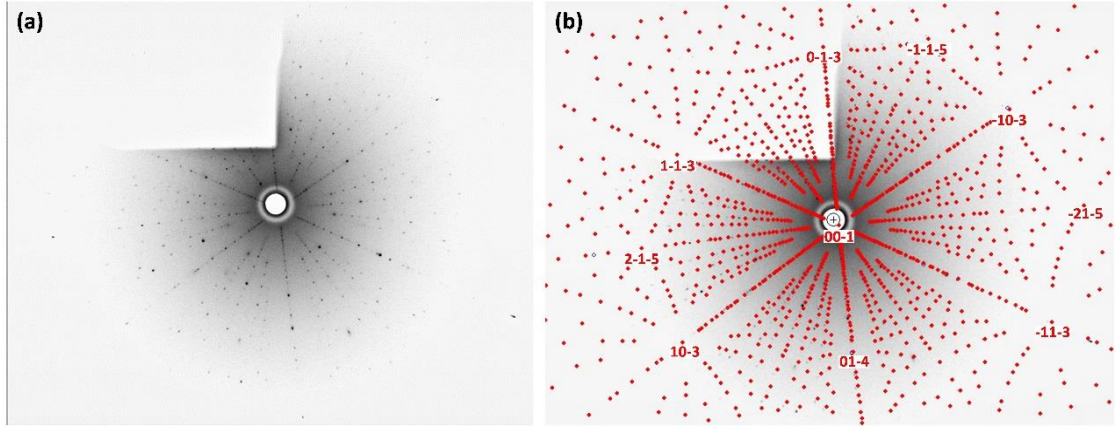

**Fig. S5.** (a) Laue diffraction pattern of the Mg$_3$Sb$_2$ single crystal. (b) The diffraction pattern can be indexed based on $P\bar{3}m1$ space group, as superposed with a theoretically simulated pattern. The corresponding diffraction spots locate exactly up to the indices.

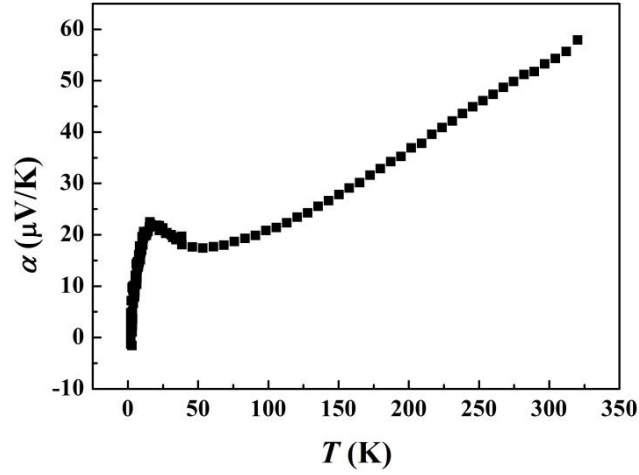

**Fig. S6.** Temperature dependence of Seebeck coefficient for the Mg$_3$Bi$_2$ single crystal along $b$ axis.

**Modeling analysis for $C_p$**

The low temperature specific heat capacity $C_p$ was fitted by using the Debye-Einstein model:[1]

$$C_p = \delta T + 9N_A k_B \left(\frac{T}{\theta_D}\right)^3 \int_0^{x_D} \frac{x^4 e^x}{(e^x - 1)^2} dx + \sum_{i=1}^{n} A_i\left(\theta_{E_i}\right)^2 (T^2)^{-\frac{3}{2}} \frac{e^{\frac{\theta_{E_i}}{T}}}{\left(e^{\frac{\theta_{E_i}}{T}} - 1\right)^2},$$

$$x = \hbar\omega/k_B T \tag{S1}$$

Where $\delta T$ denotes the heat capacity contributed by electrons, the second term denotes the part described by the Debye model of specific heat capacity and the third term denotes the total contributions of each Einstein oscillators with Einstein temperature, $\theta_{Ei}$ ($n$ is the number of the Einstein oscillators). And $\delta$, Debye temperature $\theta_D$, and $A_i$ are the fitting parameters.

As the Einstein terms of heat capacity has little influence at very low temperature ($T \leq \theta_D/10$) and the Debye terms could be simplified into the well-known $T^3$ relation, $C_v$ has the form:

$$C_p = \delta T + \frac{12}{5}\pi^4 n N_A k_B \left(\frac{T}{\theta_D}\right)^3 \tag{S2}$$

By fitting experimental data below certain temperatures ($T^2 < 10$ K$^2$ for Mg$_3$Bi$_2$ and $T^2 < 20$ K$^2$ for Mg$_3$Sb$_2$), $\delta$ and $\theta_D$ were obtained.

Then, Einstein terms were added to fit the data with broader temperature range. It was found that one Einstein oscillators are adequate to realize good fits with the experimental data. All the above mentioned fitting parameters and fitting results could be seen in Table S1 and Fig. S7.

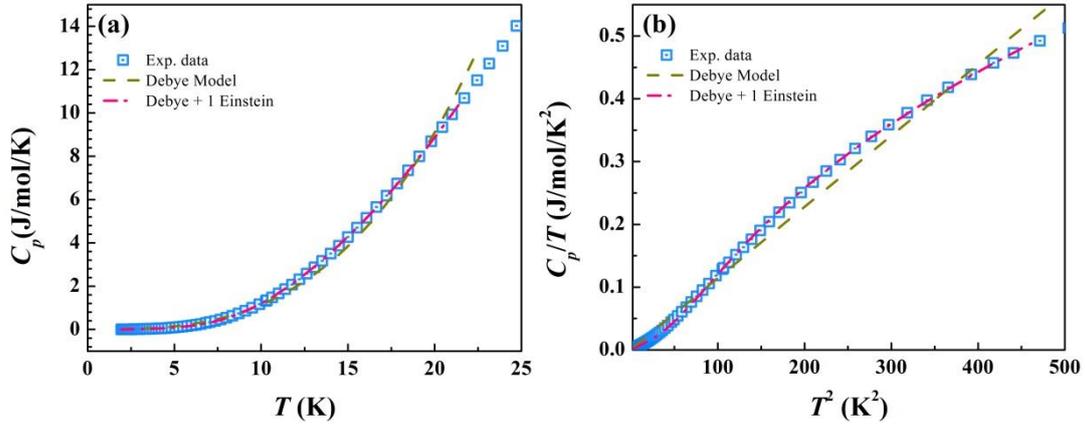

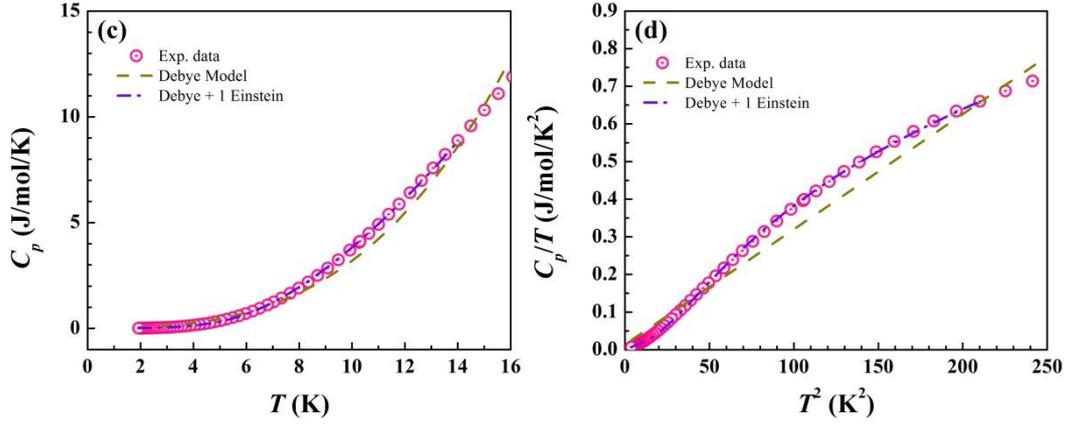

**Fig. S7.** Temperature dependence of specific heat capacity of (a, b) $Mg_3Sb_2$ and (c, d) $Mg_3Bi_2$.

**Table S1.** Fitting parameters for the specific heat capacity

|  | $\delta$ (mJ/mol/K$^2$) | $\theta_D$ (K) | $A_1$ (J/mol/K) | $\theta_{E1}$ (K) | $f_{E1}$ (cm$^{-1}$) |
|---|---|---|---|---|---|
| **$Mg_3Sb_2$** | 0.084 | 241 | 7.31 | 64 | 44.2 |
| **$Mg_3Bi_2$** | 0.133 | 175 | 8.43 | 44 | 30.8 |

**Thermal conductivity analysis based on the modified Debye-Callaway model**

The experimental lattice thermal conductivity of the single crystals were analyzed by using the modified Debye-Callaway model proposed by Morelli *et al.*,[2] in which longitude and transverse acoustic phonon modes are calculated individually and the fitting parameters for Normal and Umklapp process are transformed to the Grüneisen parameters of each phonon modes. Calculation details go as follows.

The lattice thermal conductivity $\kappa$ is expressed as

$$\kappa = \kappa_L + 2\kappa_T \tag{S3}$$

where $\kappa_L$ and $\kappa_T$ are the lattice thermal conductivity contributed by longitude and transverse acoustic phonon modes respectively. And thermal conductivity of each modes could be expressed as a uniform expression:[3, 4]

$$\kappa_i = \kappa_{i1} + \kappa_{i2}, (i = L, T) \tag{S4}$$

where $\kappa_{i1}$ represents the overall thermal conductivity for the longitude or transverse mode considering all the scattering process, and $\kappa_{i2}$ could be understood as a correction term due to the non-resistive three phonon Normal process (N). And $\kappa_{i1}$ and $\kappa_{i2}$ could be expressed as:

$$\kappa_{i1} = \frac{1}{3} C_i T^3 \int_0^{\theta_i/T} \frac{\tau_C^i x^4 e^x}{(e^x - 1)^2} dx \tag{S5a}$$

$$\kappa_{i2} = \frac{1}{3} C_i T^3 \frac{\left[\int_0^{\theta_i/T} \frac{\tau_C^i x^4 e^x}{\tau_N^i (e^x-1)^2} dx\right]^2}{\int_0^{\theta_i/T} \frac{\tau_C^i x^4 e^x}{\tau_N^i \tau_R^i (e^x-1)^2} dx}, \quad C_i = \frac{k_B^4}{2\pi^2 \hbar^3 v_i} \tag{S5b}$$

Where $x = \hbar\omega/k_B T$ is the reduced phonon energy, $\hbar$ is the reduced Planck constant, $k_B$ is the Boltzmann constant, $\omega$ is the phonon frequency, $\theta_i$ and $v_i$ are the Debye temperature and the sound velocity for specific phonon mode, $\tau_c^i$ is the combined relaxation time for phonon scattering processes, calculated by $(\tau_C^i)^{-1} = (\tau_R^i)^{-1} + (\tau_N^i)^{-1}$. $\tau_R^i$ represents the relaxation time for all the phonon resistive scattering process and $\tau_N^i$ represents the relaxation time for phonon non-resistive scattering process (N process). For the phonon resistive scattering process, the Umklapp process scattering (U), boundary scattering (B), point defect scattering (PD), and resonant scattering (PR) are taken into consideration. Therefore,

$$(\tau_R^i)^{-1} = \sum_j (\tau_j^i)^{-1}, \quad j = U, B, PD, PR \tag{S6}$$

For the relaxation time of U-process, the expression proposed by Slack *et al.* was adopted:[5]

$$(\tau_U^i)^{-1} = B_U^i \left(\frac{k_B}{\hbar}\right)^2 x^2 T^3 e^{-\theta_i/3T}, \quad B_U^i \approx \frac{\hbar \gamma_i^2}{M v_i^2 \theta_i} \tag{S7}$$

Where $M$ is average atomic mass and $\gamma_i$ is the Grüneisen parameters of each phonon modes (set as fitting parameters).

For point defect scattering,[6, 7]

$$(\tau_{PD}^i)^{-1} = \frac{V k_B^4 \Gamma}{4\pi \hbar^4 v_i^3} x^4 T^4 \tag{S8}$$

Where $V$ is average atomic volume, and $\Gamma$ is the disorder parameter due to the mass fluctuation and strain fluctuation of the point defects which was set as a fitting parameter for the sake of simplicity.

For boundary scattering,

$$(\tau_B^i)^{-1} = \frac{v_i}{L} \tag{S9}$$

Where $L$ is the critical length of boundary scattering, also set as a fitting parameter for better accordance in very low temperature range.

For resonant phonon scattering,[8]

$$(\tau_{PR}^i)^{-1} = \frac{C_1 \omega^2}{(\omega_{0,1}^2 - \omega^2)^2} \tag{S10}$$

Where $\omega_{o,1}$ is the resonance frequency of the oscillator and $C_1$ is the fitting parameters.

For N-process which was the least studied, we simply adopted the expression from the reference:[2]

$$(\tau_N^L)^{-1} = B_N^L \left(\frac{k_B}{\hbar}\right)^2 x^2 T^5, \quad B_N^L \approx \frac{k_B^3 \gamma_L^2 V}{M \hbar^2 v_L^5} \tag{S11a}$$

$$(\tau_N^T)^{-1} = B_N^T \left(\frac{k_B}{\hbar}\right) x T^5, \quad B_N^T \approx \frac{k_B^4 \gamma_T^2 V}{M \hbar^3 v_T^5} \tag{S11b}$$

The parameters used in the calculations are listed in Table S2.

**Table S2.** Physical parameters used in the modeling analysis

|  | $Mg_3Sb_2$ | $Mg_3Bi_2$ |
|---|---|---|
| $M$ /$10^{-25}$ kg | 1.05 | 1.63 |
| $V$ /$10^{-29}$ m$^{-3}$ | 2.67 | 2.88 |
| $v_L$ /m | 3794 | 3120 |
| $v_T$ /m | 1984 | 1522 |
| $\theta_L$ /K | 378 | 303 |
| $\theta_T$ /K | 198 | 148 |
| $\omega_{o,1}$ /THz | 8.33 | 5.80 |

Note: the sound velocity is from the experimental values[9]. The Debye temperature was calculated according to $\theta i = \hbar v i (6\pi^2 n)^{1/3}/k_B$, where $n$ is the atom number in a unit volume. The resonant frequency is determined by the specific heat capacity.

**Table S3.** Fitting parameters for the modeling analysis

|  | $Mg_3Sb_2$ | $Mg_3Bi_2$ |
|---|---|---|
| $L$ /mm | 0.661 | 0.173 |
| $\gamma_L$ | 4.33 | 6.49 |
| $\gamma_T$ | 2.22 | 2.22 |
| $\Gamma$ | 0.0152 | 0.0109 |
| $C_1$ /s$^{-3}$ | 3.68×10$^{34}$ | 5.27×10$^{34}$ |